\documentclass[%
 reprint,
 amsmath,amssymb,
 aps,
]{revtex4-1}

\usepackage{graphicx}
\usepackage{dcolumn}
\usepackage{bm}

\begin{document}

\preprint{APS/123-QED}

\title{The nuclear polaron beyond the mean-field approximation}

\author{D. Scalbert}%
 \affiliation{Laboratoire Charles Coulomb (L2C), UMR 5221 CNRS-Universit\'{e} de Montpellier, Montpellier, FR-34095, France}

\date{\today}

\begin{abstract}
In III-V semiconductors it was shown theoretically that under optical cooling the nuclear spin polaron bound to neutral donors would form below some critical nuclear spin temperature $T_C$ [I. A. Merkulov, Phys. Solid State \textbf{40}, 930 (1998)]. The predicted critical behavior is a direct consequence of the use of the mean-field approximation. It is known however that in any finite size system a critical behavior must be absent. Here we develop a model of the optically cooled nuclear polaron, which goes beyond the mean-field approximation. An expression of the generalized free energy of the optically cooled nuclear polaron, valid for a finite, albeit large, number of spins, is derived. This model permits to describe the continuous transition from the fluctuation dominated regime to the saturation regime, as the nuclear spin temperature decreases. It is shown that due to the finite number of nuclear spins involved in the polaron, the critical effects close to $T_C$ are smoothed by the spin fluctuations. Particularly, instead of a divergence, the nuclear spin fluctuations exhibit a sharp peak at $T_C$, before being depressed well below $T_C$. Interestingly, the formation of the nuclear polaron can, in certain conditions, boost the nuclear polarization beyond the value obtained solely by optical pumping.  Finally, we suggest that the nuclear polaron could be detected by spin noise spectroscopy or via its superparamagnetic behavior.


\end{abstract}

\maketitle

\section{\label{sec:Intro} Introduction}

Nuclear spins in semiconductors are actively studied because they exhibit a rich physics mainly related to the hyperfine interaction with the carriers \cite{OpticalOrientation,DyakonovBook,Urbaszek2013}.  In addition they  are of great importance for  applications, such as quantum information processing, or for boosting the NMR signal especially in biological systems. Regarding quantum information processing, on one hand they are considered as promising qbits because of their long spin coherence times \cite{Morton2008}. On the other hand they form a fluctuating spin bath which contributes to the decoherence of localized electrons  \cite{Merkulov2002,Khaetskii2002}, and also limits the generation of indistinguishable photons in quantum dots \cite{Malein2016}. Different methods have been proposed to reduce or eliminate the effect of the nuclear field fluctuations on the electron spin dephasing time among these, dynamical decoupling \cite{deLange2010}, feedback loop that controls the nuclear spin bath \cite{Bluhm2010}, or polarization of nuclei close to saturation in order to reduce the nuclear field fluctuations \cite{Taylor2003,CHrist2007}. This last option could be accomplished by optical spin-pumping, where the electron spin polarization is transferred to the nuclei via the hyperfine interaction. Remarkable progress has been made in this direction, but the highest nuclear spin polarization achieved (65\% in QDs \cite{Chekhovich2010,Ulhaq2016}) is still insufficient for a sizable reduction of fluctuations.

Almost complete nuclear spin polarization could be reached in case of nuclear polaron formation. The conditions for nuclear polaron formation after optical cooling of the nuclei has been discussed by Merkulov \cite{Merkulov1998}. Merkulov has shown that the nuclear polaron will form if the product of the nuclear spin temperature $T_n$ and the lattice temperature $T_e$ is positive, and less than some critical value. For the shallow donor in GaAs the critical nuclear spin temperature will be  $T_C\sim 10^{-7}\text{ K}$ if the lattice is at 4~K. Although quite low this temperature can in principle be reached by optical spin pumping followed by adiabatic demagnetization. In GaN, where the Bohr radius is about 4 times smaller than in GaAs (number of nuclear spins $\sim 10^4$) and the nuclear field at saturation is 10 times less \cite{Wang2014}, the critical temperature will be about 10 times larger. Quantum dots containing a resident electron can also in principle exhibit polaron effect. However in this case the strong quadrupolar interactions prohibit the establishing of a temperature within the nuclear spin system \cite{Maletinsky2009}.

Here we develop a model of the optically cooled nuclear polaron which goes beyond the mean-field approximation of Merkulov's model, and which takes into account the thermodynamic nuclear spin fluctuations. This is mandatory to describe a finite size system, with $10^4-10^6$ nuclear spins, and will allow us to assess the effect of these fluctuations on the electron spin dephasing time. We will see that the spin fluctuations exhibit a peak at $T_C$, and then drop rapidly below $T_C$.

In the next Section we briefly recall the main lines of Merkulov's model. In Section III we derive the expression of the generalized free-energy adapted to an optically cooled nuclear polaron, where the nuclear and electronic spins are at two different temperatures. In Section IV we use this expression to calculate the total spin of polaron and its fluctuations, and show the progressive emergence of the critical behavior as the size of the polaron grows. Then in Section V we examine to which extent the finite heat capacity of the nuclear thermostat constituted by the nuclear spin-spin interactions limits the polaron formation. We demonstrate in Section VI the superparamagnetic properties of the nuclear polaron, which can be used as a signature of polaron formation, before concluding.


\section{Merkulov's model}
In Merkulov's model the hyperfine interaction between the donor electron spin and the nuclear spins is treated within the mean-field approximation (MFA).
The hamiltonian describing the hyperfine contact interaction takes the form \cite{OpticalOrientation}
\begin{equation}\label{Eq_Hhyp}
  H=A v_0\mathbf{s}_e\cdot \sum_{n=1}^{N} |\psi(\mathbf{R}_n)|^2\mathbf{I}_n,
\end{equation}
where $A$ is the hyperfine constant, $v_0$ the volume of the unit cell, $\mathbf{s}_e$ the electron spin, $\mathbf{I}_n$ nuclear spin at lattice site $\mathbf{R}_n$, and $\psi(\mathbf{r})$ is the electron envelop function.

In the framework of MFA the nuclear spins and the electron spin are mutually polarized in their respective hyperfine fields. This situation is described by the set of coupled equations
\begin{equation}\label{Eq:electronfield}
  \langle I_n\rangle=\frac{1}{3}\beta_n I(I+I)Av_0|\psi(\mathbf{R}_n)|^2\langle s_e\rangle,
\end{equation}
\begin{equation}\label{Eq:nuclearfield}
   \langle s_e\rangle=\frac{1}{3}\beta_e s(s+I)Av_0\sum_{n}|\psi(\mathbf{R}_n)|^2\langle I_n\rangle,
\end{equation}
$\beta_e$ and $\beta_n$ are the inverse electron and nuclear spin temperatures. For $\beta_e\beta_n>0$ and at low enough electron and nuclear spin temperatures these equations admit a non-trivial solution corresponding to the polaron formation. The  critical nuclear spin temperature at which the polaron forms is given by
\begin{equation}\label{Eq:TC}
  T_C=\beta_e\frac{A^2 I(I+1)}{12 k_B}v_0^2 \sum_{n}|\psi(\mathbf{R}_n)|^4.
\end{equation}

For an hydrogenoid wavefunction $\psi(r)=1/\sqrt{\pi a_B^3}\exp(-r/a_B)$ this expression reduces to
\begin{equation}\label{Eq:TChyd}
  T_C=\beta_e\frac{A^2 I(I+1)}{48\pi k_B}\frac{v_0}{a_B^3}.
\end{equation}

\section{Generalized free-energy}\label{Sec:Free-energy}
We seek an expression of the out-of-equilibrium, or generalized, free-energy valid when electron and nuclear spin temperatures are not equal, and from which the probability of any spin configuration can be calculated. This is necessary to account for the finite size of the polaron, and look for the influence of the thermodynamic fluctuations on the phase transition predicted by the MFA.

\begin{figure}
  \centering
  \includegraphics[width=8cm]{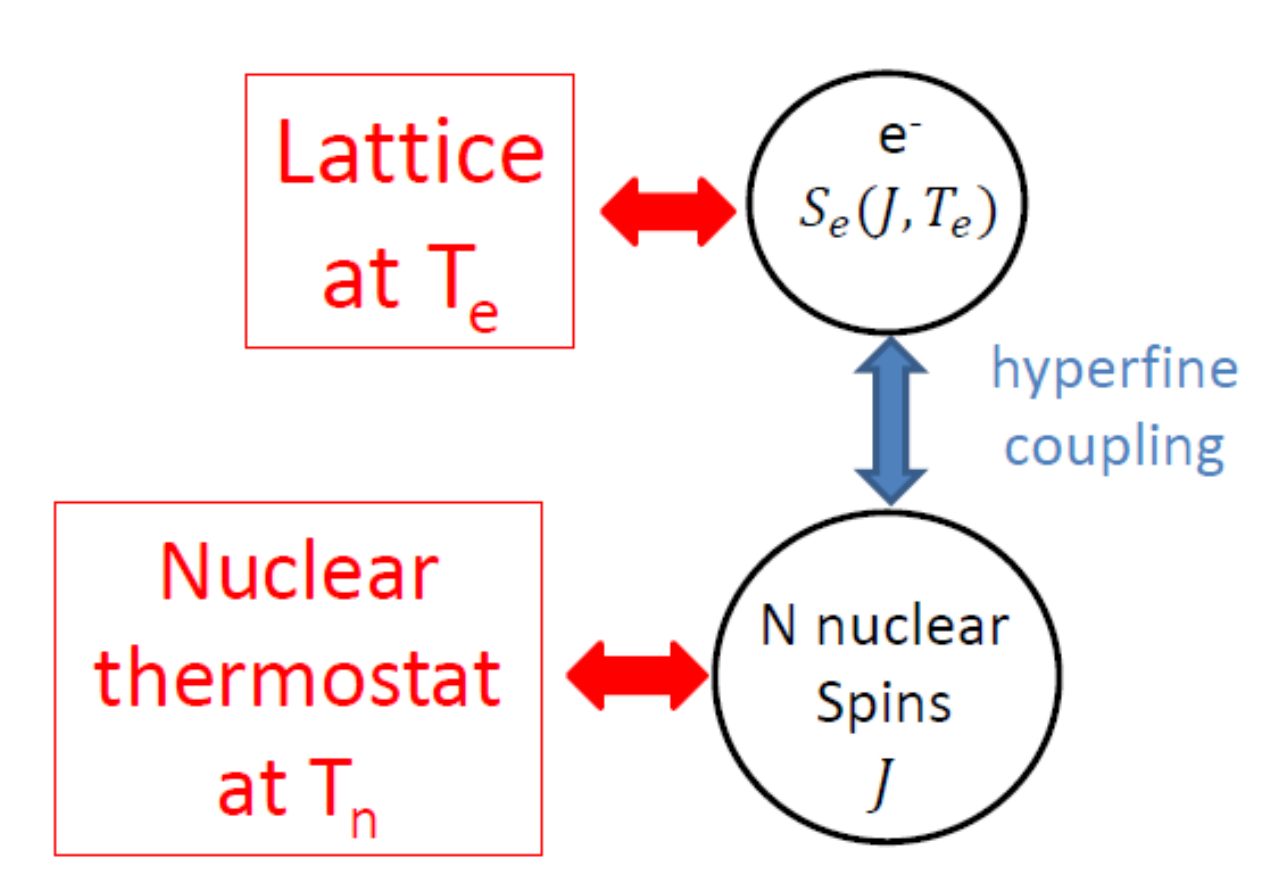}
  \caption{Schematics of the interacting electron-nuclear spin system forming the nuclear polaron, in which the electron spin is thermalized to the lattice, whereas the nuclear spins are thermalized to another thermostat formed by the nuclear spins. }\label{Fig:polaron scheme}
\end{figure}

This system is not stricto sensu in thermodynamic equilibrium, since the two subsystems, the electron spin and the nuclear spins, are in thermal contact with two thermostats at different temperatures. The electron spin thermostat is the lattice, while the nuclear thermostat is constituted by the nuclear spin-spin interactions, and by the nuclear spins outside the polaron, which are in thermal contact via spin diffusion (see Fig.~\ref{Fig:polaron scheme}). We assume for the time being that this constitutes a good thermostat. As the electron and the nuclear spins are coupled by the hyperfine interaction, heat can flow across the system in order to restore thermal equilibrium. However this heat flow can be small enough to consider the system to be in quasi equilibrium, so that one can still use the laws of equilibrium thermodynamics.

Therefore we consider the electron spin and the nuclear spins as two distinct subsystems coupled to different thermostats, and weakly interacting via the hyperfine hamiltonian (Fig.~\ref{Fig:polaron scheme}).

To make calculations tractable we consider the box model with a step wave function $\psi(r)=(\frac{4}{3}\pi a_0^3)^{-1/2}\Theta(a_0-r)$. As crude as it may seem this approximation turns out to give results equivalent to those obtained with a realistic hydrogenoid wave function provided one choose the box radius such that $a_0=6^{1/3}a_B$, which amounts to keep the same value for the integral $\int |\psi(r)|^4 d^3r$ \cite{DMS} . The hyperfine hamiltonian thus becomes
\begin{equation}\label{boxmodel}
  H=a \textbf{s}_e\cdot \textbf{J},
\end{equation}

with $a=A v_0/V$, $V=4\pi a_0^3/3$, and $J=\sum_{n=1}^{N} \textbf{I}_n$.

Since the electron spin evolves much rapidly than the nuclear moment, at each moment it follows the nuclear polarization $J$. Therefore its polarization is simply given by the thermal value $s_e(J,T_e)$ in the quasi-static instantaneous nuclear moment $J$. Conversely, the nuclear spins are subjected to the average field created by the electron, which acts as an external field. The hyperfine interaction energy becomes
\begin{equation}\label{Eq_Heff}
  E(J,T_e)=as_e(J,T_e)J.
\end{equation}

We must now evaluate the probability $P(J)$ of each macrostate $J$. Since the nuclear spins are in equilibrium with a thermostat this probability is given by the Boltzmann-Gibbs distribution $P(J)=g(J)\exp(-\beta_n E_n(J))$, where $g(J)$ is the degeneracy of the macrostate $J$ including the orientational degeneracy. One must be careful in the definition of $E_n(J)$. In order to properly define $E_n(J)$ one has to calculate the amount of heat transfer between the hyperfine energy reservoir and the nuclear thermostat accompanying a variation of the macrostate from $J$ to $J+\delta J$. Such a transformation will necessarily also involve a heat transfer to the lattice, since the electron spin reacts instantaneously to the variation of nuclear polarization. This heat transfer does not contribute to the establishing of the Boltzmann-Gibbs distribution for the nuclear spins and must be removed from the total energy variation given by
\begin{equation}\label{}
  \delta E=as_e(J,T_e)\delta J+a\frac{\partial s_e(J,T_e)}{\partial J}J\delta J.
\end{equation}
Only the first term corresponding to the energy variation in the constant field of the electron must be retained. The second term corresponds to a variation of the electron spin polarization, which involves the heat transfer with the lattice. By integration one gets the expression of the energy which is involved in the energy transfer between the nuclear system and its thermostat
\begin{equation}\label{Eq_En}
  E_n(J)=-\frac{1}{\beta_e}\log\left[\cosh(\frac{1}{2}\beta_e aJ)\right].
\end{equation}
From this we deduce the expression of the Boltzmann-Gibbs distribution in the general case when $\beta_e\neq\beta_n$

\begin{equation}\label{Eq_Boltzmann-Gibbs}
  P(J,\beta_e,\beta_n)=C g(J)[\cosh(\frac{1}{2}\beta_e a J)]^{\beta_n/\beta_e},
\end{equation}
\medskip
where $C$ is a normalization constant for the probability distribution $P(J,\beta_e,\beta_n)$.

Next we must evaluate $g(J)$. In case of spins one-halves an exact expression has been derived by Kozlov \cite{Kozlov2007}. For spins larger than one-half and for a small number of spins $g(J)$ can be computed iteratively, but for large $N$ it is more efficient to use an approximate expression. Such an expression can be easily obtained by considering the generalized free energy of N independent spins in an external field. The generalized free energy is a function of the total spin projection $J_z$ along the magnetic field $\omega$
\begin{equation}\label{free energy}
  F(J_z)=-\omega J_z-\frac{1}{\beta}\log(g_z(J_z)),
\end{equation}
where $g_z(J_z)$ represents how much times the resultant projection $J_z$ can be obtained by adding $N$ spins. Noting that for large $N$ the thermodynamic average of $J_z$ is obtained by minimization of $F(J_z)$ at any constant inverse temperature $\beta$ one obtains
\begin{equation}\label{Fderivative}
  \frac{d}{dJ_z}[\log(g_z(J_z))]=-\beta\omega,
\end{equation}
where $\beta$ and $J_z$ are related through the Brillouin function by $J_z=NIB_I(\omega\beta I)$. Using $g_z(NS)=1$ by integration of Eq.~ (\ref{Fderivative}) one gets
 \begin{equation}\label{}
  g_z(J_z)=\exp \left[N\int_{|p|}^{1}B_I^{-1}(y)dy\right],
\end{equation}
where $p=J_z/NS$ is the spin polarization. The entropy per spin is given by
\begin{equation}\label{eq:entropy}
\begin{split}
     & \frac{S}{N}=k_B\int_{|p|}^{1}B_I^{-1}(y)dy \\
     & =k_B\left[\log(Z(B_I^{-1}(|p|)))-pB_I^{-1}(p)\right],
\end{split}
\end{equation}
where the integration is easily carried out by a variable change and by using the relation $B_I(x)=[\log(Z_I(x)]'$, where $Z_I(x)=\sinh(\frac{2I+1}{2I}x)/\sinh(\frac{x}{2I})$ is the partition function for a spin $I$. Note that Eq.~(\ref{eq:entropy}) can also be obtained by the steepest-descent method  \cite{Kubo,Petukhov2007}.

\begin{figure}
  \centering
  \includegraphics[width=8 cm]{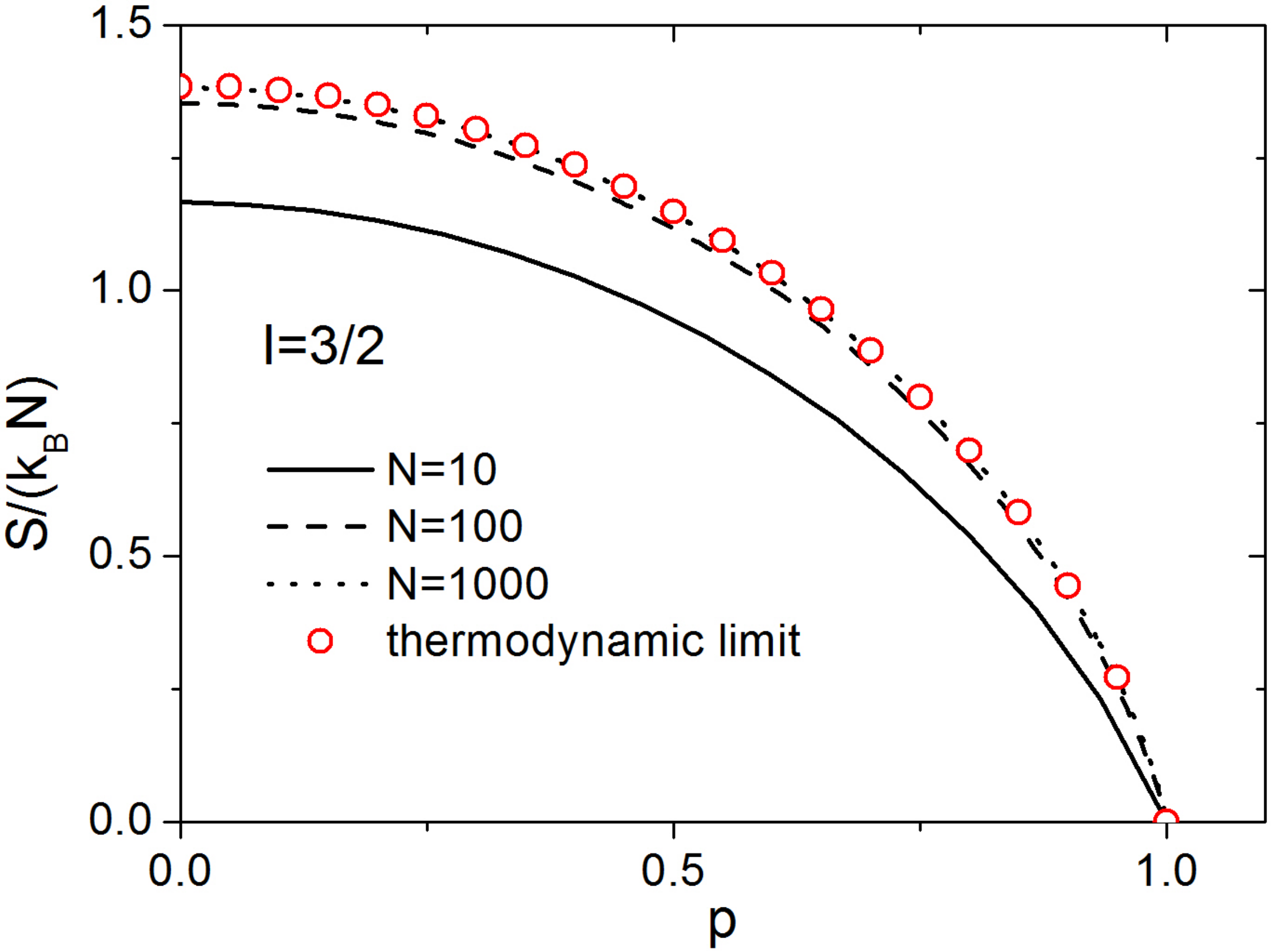}
  \caption{Solid, dashed, and dotted lines represent the exact entropy per spin, in units of $k_B$, for $N$ independent spins, calculated quantum mechanically by iterative method. The red circles represent the thermodynamic limit $N\rightarrow \infty$ calculated with Eq.~(\ref{eq:entropy}). The agreement is already satisfactory for $N=100$, and becomes quite good for $N=1000$. }\label{fig:entropy}
\end{figure}

Figure \ref{fig:entropy} shows that already for 1000 spins the entropy is very well approximated by Eq.~(\ref{eq:entropy}), i.e. the thermodynamic limit is reached.

\medskip

 The degeneracy $g(J)$ is then deduced by using $g(J)=-(2J+1)g'_z(J)$. One finds
\begin{equation}\label{Eq:gJ}
\begin{split}
   g(J)= &\frac{2J+1}{I}B_I^{-1}\left(\frac{J}{NI}\right)\times \\
     & \exp\left[N\log(Z_I(B_I^{-1}(\frac{J}{NI})-\frac{J}{I}B_I^{-1}(\frac{J}{NI})\right].
\end{split}
\end{equation}

Combining the results of Eq.~(\ref{Eq_En})-(\ref{Eq:gJ}) one gets the generalized free energy for the optically cooled nuclear polaron
\begin{widetext}
\begin{equation}\label{Eq:free energy}
  F(J,\beta_e,\beta_n)=-\frac{1}{\beta_e}\log\left[\cosh(\frac{1}{2}\beta_e aJ)\right]-\frac{1}{\beta_n}\left\{ \log\left[ (2J+1)B_I^{-1}(\frac{J}{NI})\right] +N \log \left[Z_I(B_I^{-1}(\frac{J}{NI}))\right]-\frac{J}{I}B_I^{-1}(\frac{J}{NI}) \right\}.
\end{equation}
\end{widetext}

At high nuclear temperature ($\beta_n\rightarrow0$) the right-hand side of Eq.~(\ref{Eq:free energy}) can be developed up to second order in $J/NI$. We obtain
\begin{widetext}
\begin{equation}\label{Eq:GaussianApprox}
 \lim_{\beta_n \rightarrow 0} \exp\left[-\beta_nF(J,\beta_e,\beta_n)\right]\simeq \frac{6J^2 (2I+1)^N}{NI(I+1)}\exp\left[\frac{-3J^2}{2NI(I+1)}\right].
\end{equation}
\end{widetext}
The gaussian approximation is recovered. In addition in the case $(\beta_e=\beta_n) \rightarrow 0$, and for $a_0=6^{1/3}a_B$, Eq.~(\ref{Eq:free energy}) becomes equivalent to the Dietl-Spalek's theory of the magnetic polaron with thermodynamic fluctuations included \cite{Dietl1982,Dietl1983}. Namely, the distribution of electron spin splittings are the same in both cases. This shows that the box model adequately describes this basic property of the nuclear polaron, not only in the low temperature regime, where the MFA is valid, but also above the critical temperature.

\section{Emergence of the critical behavior}

Let us define the normalized total spin of the polaron as $j= J/NI$ and $j_T$ its thermodynamic average.   At high nuclear temperature $j_T$ tends to $j_\infty=[\frac{8}{3\pi}\frac{I+1}{NI}]^{1/2}$.

At arbitrary nuclear temperature $j_T$ can be calculated using

\begin{equation}\label{}
  j_T(\beta_e,\beta_n)=\frac{1}{NI}\frac{\int_{0}^{\infty}Je^{-\beta_nF(J,\beta_e,\beta_n)}dJ}{\int_{0}^{\infty}e^{-\beta_nF(J,\beta_e,\beta_n)}dJ}.
\end{equation}

\begin{figure}
  \centering
  \includegraphics[width=8 cm]{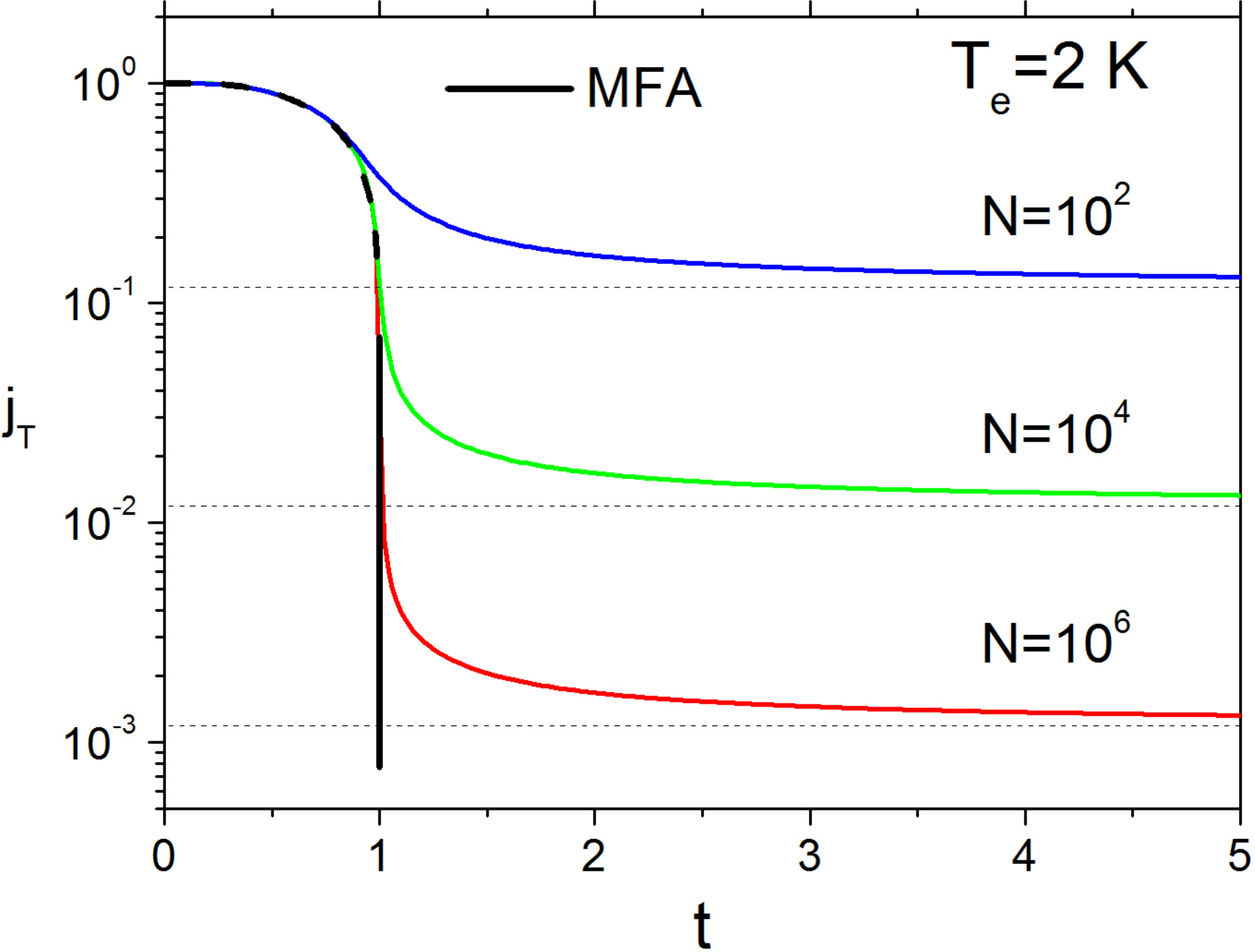}
  \caption{Nuclear polaron spin $j_T$ versus the reduced nuclear spin temperature calculated within the MFA (dark curve), and within the model beyond the MFA (colored curves). The different values of $N$ correspond to equivalent Bohr radii 0.45, 2.1, and 9.65~nm for increasing $N$. The dotted lines correspond to the high temperature limit $j_\infty$, when the polaron spin is entirely determined by the spin fluctuations.}\label{Fig:NuclearPolarization}
\end{figure}

Figure~\ref{Fig:NuclearPolarization} shows the evolution of $j_T$ versus the reduced nuclear spin temperature $t=T_n/T_C$. Here, and in all subsequent calculations the hyperfine constant is taken as $A=4.5\times 10^{-5} \text{ eV}$, corresponding to an average value for different nuclei of GaAs \cite{Paget1977}, and the electron spin temperature is fixed at $T_e=2 \text{ K}$. There is a quite good agreement between MFA and the present model below $T_C$. However, above $T_C$, the sharp transition predicted by MFA is smoothed by the spin fluctuations. Although the transition becomes sharper as $N$ increases, still for $N=10^6$ (the typical number of nuclei forming the donor bound polaron in GaAs) the polaron begins to form above $T_C$.
\medskip

\begin{figure}
  \centering
  \includegraphics[width=8 cm]{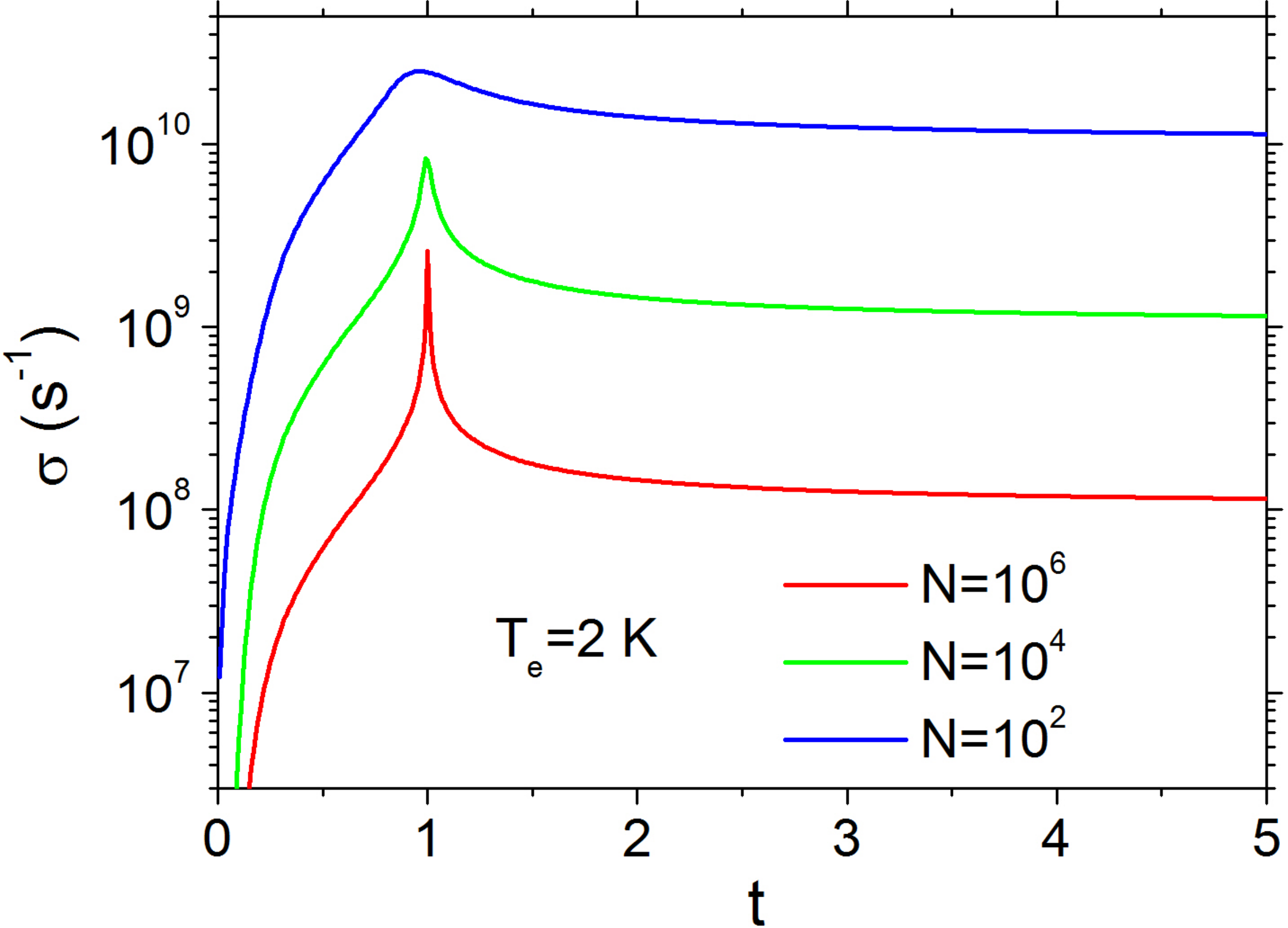}
  \caption{Standard deviation of the electron precession frequency in the polaron versus the reduced nuclear spin temperature, calculated for different $N$ as in Fig.~(\ref{Fig:NuclearPolarization}).}\label{Fig:StandardDeviation}
\end{figure}

Figure~\ref{Fig:StandardDeviation} shows the standard deviation $\sigma$ of the electron spin splitting $\omega=\frac{2A}{N\hbar}J$, which is a measure of the amplitude of thermodynamic fluctuations. As expected $\sigma$ exhibits a peak at $t=1$, which sharpens as $N$ increases. This is the signature of the progressive build up of the critical fluctuations as the size of the system increases. These fluctuations would ultimately diverge for an infinite system, when a genuine phase transition takes place.

 As the temperature is lowered below $T_C$ the fluctuations sharply decrease below the high temperature value. Correspondingly the electron spin dephasing time will be considerably enhanced in this regime.

\section{Effect of the finite heat capacity of the nuclear thermostat}

\begin{figure}
  \centering
  \includegraphics[width=8 cm]{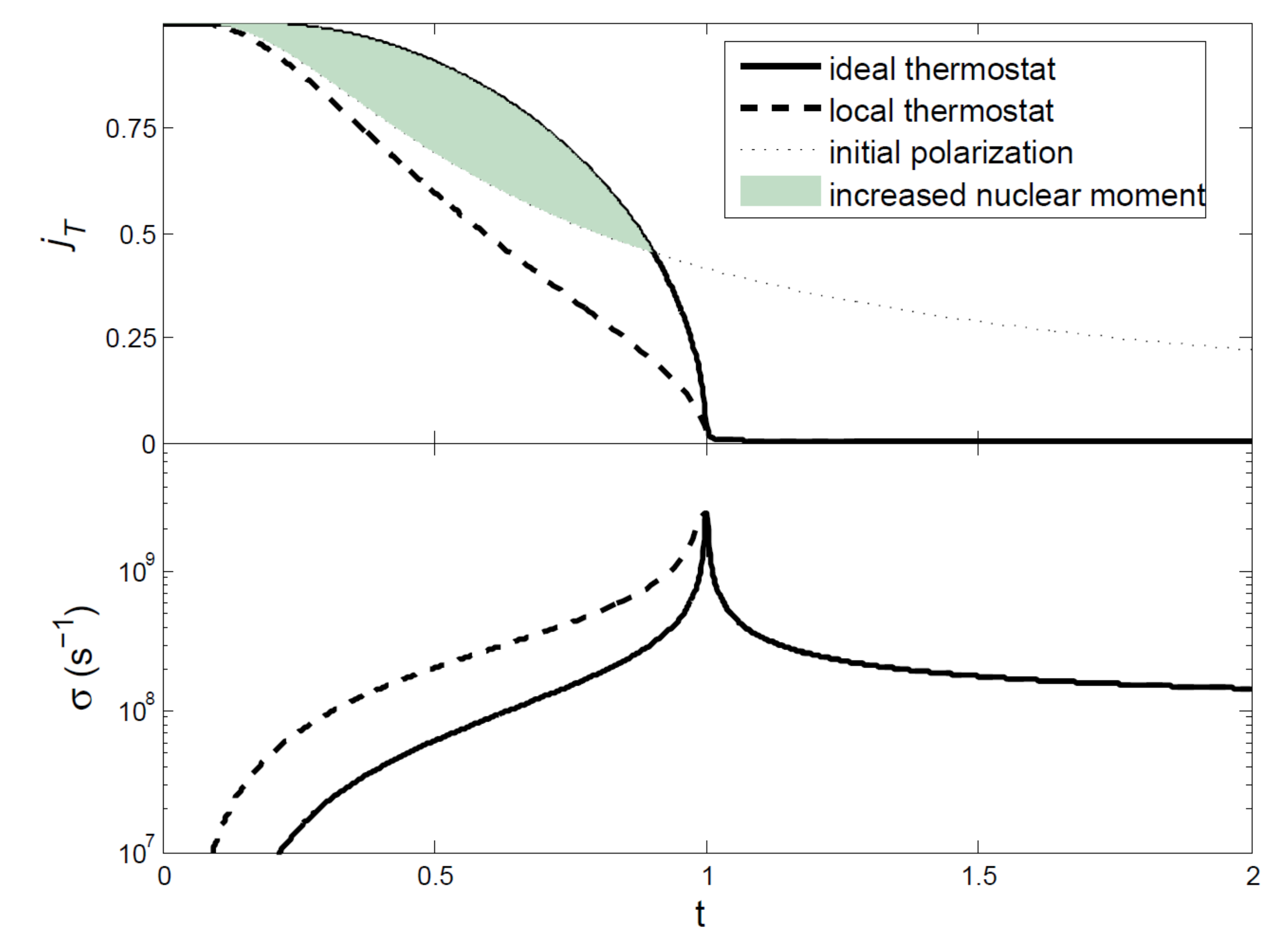}
  \caption{Effect of the finite heat capacity of the nuclear thermostat on the nuclear polaron moment $j_T$ (upper panel), and on the standard deviation of the electron spin splitting $\sigma$ (lower panel). The dotted line in the upper panel shows the initial nuclear polarization before adiabatic demagnetization. The filled area represents the values of $j_T$ for which a net gain in the nuclear spin momentum is obtained by polaron formation  ($B_L=2\text{ G}$, $N=10^6$). }\label{fig:FiniteInfinite}
\end{figure}

Until now we assumed an ideal nuclear thermostat, able to accommodate the heat dissipation, which occurs during the polaron formation. In the case of the optically cooled nuclei the polaron formation is an isentropic transformation occurring at short time scale compared to the nuclear spin-lattice relaxation. In this situation the nuclear temperature is not constant during polaron formation, but raises due to the finite heat capacity of the local thermostat constituted by the nuclear spin-spin interactions. The final nuclear temperature after polaron formation is then given by \cite{Abragam1958}
\begin{equation}\label{Eq:FinalTemp}
  \beta_n^f=\frac{B_L}{(B_L^2+B_P^2)^{1/2}}\beta_n,
\end{equation}
where $B_L$ is the local nuclear field, $B_P$ is the Knight field once the nuclear polaron is formed, and $\beta_n$ is the inverse nuclear temperature before polaron formation. The polaron will thus form as if the heat capacity of the nuclear thermostat was infinite but with a higher temperature defined by $\beta_n^f$. Hence, the normalized polaron spin is implicitly defined by Eq.~(\ref{Eq:FinalTemp}) plus the two following equations
\begin{equation}\label{Eq:Pfinite}
  j_\text{finite}(\beta_e,\beta_n)=j_T(\beta_e,\beta_n^f),
\end{equation}
\begin{equation}\label{}
  B_P=\frac{a}{\hbar}\langle s_e(J,\beta_e)\rangle_{\beta_e,\beta_n^f},
\end{equation}
$j_\text{finite}$ will be reduced with respect to $j$ as soon as $B_P>B_L$, which happens when $j$ exceeds $4B_L/(NI\beta_ea^2)\simeq 0.02$, that is below $T_C$ (see dashed line in Fig.~\ref{fig:FiniteInfinite}).
\medskip

Let us remind that the nuclei located outside the polaron are also cooled down by spin diffusion during optical pumping \cite{Paget1982}. Hence, the heat dissipated during the polaron formation may partly be expelled from the polaron core to the outside by diffusion, as long as it is not prohibited by the formation of a diffusion barrier. This will favor the polaron formation, so that $j_T$ will lie somewhere between the dashed and solid lines.

It is interesting to examine by how much the nuclear polaron effect can increase the initial nuclear polarization obtained by optical pumping under a magnetic field $B_{op}>>B_L$. Let $\beta_{n,op}$ be the corresponding nuclear temperature. After demagnetization the nuclear temperature will be $\beta_n=\beta_{n,op}B_{op}/B_L$. Hence, the initial nuclear polarization is given by the Brillouin function for spin $I$ and argument $\beta_n \hbar\gamma_n B_L I$. This polarization is shown as a dotted line in Fig.~\ref{fig:FiniteInfinite}. The filled area represents the region where a net gain in the nuclear polarization is obtained by nuclear polaron formation. Note however that the reduction of $j_T$ due to limited heat capacity of the spin-spin interactions depends on the exact form of the wave function. Here we assumed a constant average Knight field, while with a realistic wave function it will vary with the distance from the donor center, hence the final temperature after demagnetization given in Eq.~(\ref{Eq:FinalTemp}) must also vary.

\section{Superparamagnetism}\label{Sec:Superparamagnetism}

Magnetic polarons are known to exhibit superparamagnetism since their large magnetic moment enables fast orientation in an external magnetic field. In order to evaluate this property for the nuclear polaron we add the electronic and nuclear Zeeman terms to the hyperfine hamiltonian. The total hamiltonian reads
\begin{equation}\label{boxmodel 2}
  H=a \textbf{s}_e\cdot \textbf{J}+\omega_e s_{ez}+\omega_n J_{z},
\end{equation}
where the magnetic field is along the z-axis. $\omega_{e,n}=\hbar\gamma_{e,n}B$ where $\gamma_{e,n}$ are the gyromagnetic ratio of the electron and nuclei respectively, and $B$ is the magnetic field.
A reasoning similar to the one developed in Section \ref{Sec:Free-energy} gives the relevant energy exchanged between nuclear spins and the nuclear thermostat
\begin{equation}\label{Eq_En 2}
  E_n(J,J_z)=-\frac{1}{\beta_e}\log\left[\cosh(\frac{1}{2}\beta_e \Omega)\right]+\omega_n J_z,
\end{equation}
where $\Omega=|a \mathbf{J}+ \mathbf{\omega}_e|$ is the modulus of the total field acting on the electron spin.

\begin{figure}
  \centering
  \includegraphics[width=8 cm]{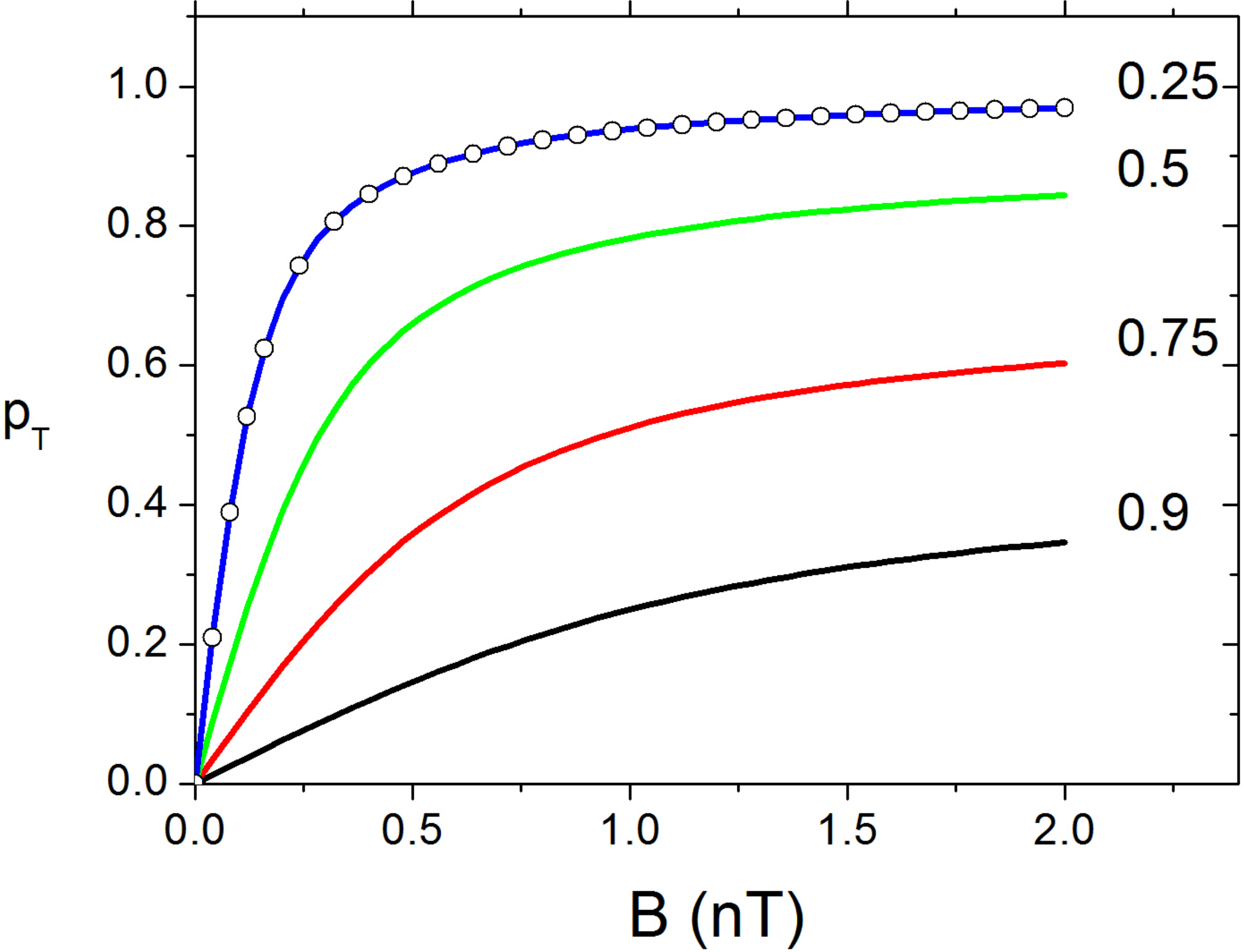}
  \caption{Nuclear polaron polarization versus magnetic field calculated for reduced nuclear spin temperatures varying from 0.25 to 0.9, $N=10^6$, $\gamma_e=-3.87\times10^{10} \text{ rad/s/T}$, $\gamma_n=6\times10^6 \text{ rad/s/T}$. The Brillouin function for spin $NI$ and $T_n=0.25T_C$ is represented by the open circles.}\label{Fig:PolaronPolarizatiobn}
\end{figure}

In evaluating the free energy one must count the number of spin configurations for a given total spin $J$ and spin projection $J_z$. This is given by $g(J)$ divided by the orientational degeneracy $2J+1$. Taking this into account the free energy of the nuclear polaron subjected to an external field is
\begin{widetext}
\begin{equation}\label{Eq:free energy 2}
  F(j,J_z,\beta_e,\beta_n)=-\frac{1}{\beta_e}\log\left[\cosh(\frac{1}{2}\beta_e \Omega)\right]+\omega_n J_z-\frac{1}{\beta_n}\left\{ \log\left[ B_I^{-1}(j)\right] +N \left [ \log \left[Z_I(B_I^{-1}(j))\right]-jB_I^{-1}(j) \right ] \right\}.
\end{equation}
\end{widetext}
We can now calculate the thermodynamic value of the polaron polarization $p_T=\langle p\rangle$. Figure \ref{Fig:PolaronPolarizatiobn} shows that below $T_C$ the polarization $p_T$ grows very rapidly in a quite small magnetic field. Note that because the applied magnetic field is much lower than the local field, $\beta_n$ remains constant in this field range. This is the gist of the nuclear polaron superparamagnetism. At low enough temperature (typically $t=0.25$) the polaron behaves as a single magnet with a very large spin $J=NI$, as evidenced by the comparison with the Brillouin function for spin $NI$.  On the contrary $p_T$ becomes quite small above $T_C$.
Let us outline that despite $\gamma_e>>\gamma_n$, the nuclear polaron orientation is governed by the nuclear spins and not by the electron, due to the large number of nuclear spins and their much lower temperature.

\section{Conclusion}

In conclusion a simplified model of the optically cooled nuclear polaron, which takes into account the thermodynamic nuclear spin fluctuations has been developed. Although the model assumes a step-like wave function its virtue is to demonstrate some important properties of the nuclear polaron. Namely, despite the very large number of nuclear spins in the polaron (typically $10^6$ for GaAs) the finite size of the system prevents a genuine phase transition.  Particularly, the polaron total spin is not zero above $T_C$ and starts to increase as the temperature is lowered from above $T_C$. Also a sharp peak in the spin fluctuations of the polaron, reminiscent of the divergence occurring for infinite systems, is predicted at $T_C$. In case of efficient spin diffusion the total spin momentum of the polaron can exceed the initial spin polarization obtained by optical pumping.

The nuclear polaron could be detected by spin noise spectroscopy, a technique recently developed in semiconductors \cite{Oestreich2005, H??bner2014}, which has proved to be efficient for measuring the nuclear field \cite{Ryzhov2015,Ryzhov2016,Smirnov2015a}. The measurements will have to be faster than the characteristic time of nuclear spins warm up \cite{[{}][{ (to appear in Phys. Rev. B.).}]Vladimirova}.  Alternatively the superparamagnetism, accompanied by a sudden rise of the donor spin polarization under a very weak magnetic field, would be a signature of the polaron formation. Although challenging the evidence of nuclear polaron in semiconductors would be one example of the various collective nuclear spin states, which can form in a deeply cooled nuclear spin system \cite{Oja1997}.

\section*{Acknowledgements}
The author enjoyed stimulating discussions with K. Kavokin.


%

\end{document}